\documentclass{SCIS2019}

\begin{document}
\ArticleType{}
\Year{2019}
\Month{}
\Vol{}
\No{}
\DOI{}
\ArtNo{}
\ReceiveDate{}
\ReviseDate{}
\AcceptDate{}
\OnlineDate{}

\title{Radiation build-up and dissipation in \\ random fiber laser}{Radiation build-up and dissipation in random fiber laser}

\author[1]{Shengtao Lin}{}
\author[1]{Zinan Wang}{{znwang@uestc.edu.cn}}
\author[1]{Jiaojiao Zhang}{}
\author[1]{Pan Wang}{}
\author[2]{Han Wu}{}
\author[1]{Yifei Qi}{}

\AuthorMark{Shengtao Lin}

\AuthorCitation{Shengtao Lin, Zinan Wang, Jiaojiao Zhang, et al}


\address[1]{Key Laboratory of Optical Fiber Sensing and Communications, \\University of Electronic Science and Technology of China, Chengdu {\rm 611731}, China}
\address[2]{College of Electronics and Information Engineering, Sichuan University, Chengdu {\rm610064}, China}

\abstract{Random fiber laser (RFL) is a complex physical system that arises from the distributed amplification and the intrinsic stochasticity of the fiber scattering. There has been widespread interest in analyzing the underlying lightwave kinetics at steady state. However, the transient state, such as the RFL build-up and dissipation, is also particularly important for unfolding lightwave interaction process. Here, we investigate for the first time the RFL dynamics at transient state, and track the RFL temporal and spectral evolution theoretically and experimentally. Particularly, with the contribution of randomly distributed feedback, the build-up of RFL shows continuous Verhulst logistic growth curves without cavity-related features, which is significantly different from the step-like growth curve of conventional fiber lasers. Furthermore, the radiation build-up duration is inversely related to the pump power, and the spectral evolution of RFL undergoes two phases from spectral density increase to spectral broadening. From steady-state to pump switch-off state, the RFL output power dissipates immediately, and the remaining Stokes lightwave from the Rayleigh scattering will gradually disappear after one round-trip. This work provides new insights into the transient dynamics features of the RFL.}


\maketitle

\section{Introduction}
Random fiber laser (RFL), generated from random distributed feedback in an amplified fiber waveguide medium \cite{Turitsyn2014}, exhibits excellent performance in high efficiency \cite{Wang2015a}, good beam quality \cite{Xu2017} and wavelength agility \cite{Zhang2017}. It has been widely used in the field of optical fiber sensing and optical imaging \cite{Churkin2015,Wang2021}, and has attracted recent attention in the high-power laser driving facility \cite{Fan2021}. Besides, RFL involves complex lightwave propagation and interaction processes with random media presenting distinctive statistical properties \cite{Li2019}, and it has become a statistical platform to research some challenging aspects of complex systems \cite{Gomes2022}.

When operating at steady-state, RFL exhibits complex lasing process. For the intensity fluctuations at steady-state, RFL shows universal statistical properties, i.e., Gaussian distribution at the prelasing regime, L\'evy behavior around threshold, and Gaussian pattern above the threshold \cite{Li2019,Lima2017}. In addition, the turbulence \cite{Gonzalez2017} and optical rogue wave \cite{Xu2020} behavior can be also observed in RFL, which contains complex interplay process. For the spectrum statistics at steady-state, the nonlinear correlation between different parts of the RFL spectra is observed through mutual information analysis \cite{Sugavanam2017}, and the underlying stimulated Brillouin scattering is revealed. Besides, the transition from the paramagnetic to the spin-glass phase in disordered magnetic systems can be also described in the RFL platform by using the Parisi overlap parameter \cite{Gomes2016,Gonzalez2018,Zhou2021}.

For the transient state like the radiation build-up and dissipation process, it is difficult to synchronously and fast track the evolution of the RFL intensity and spectrum due to the continuous and broadband characteristics of RFL lightwave \cite{Sugavanam2016}, and there is no research on the RFL transient state to the best of our knowledge.  However, analyzing the build-up and dissipation processes of the complex physical system is critical for unveiling the underlying physical mechanism. For instance, covert transient coherent multi-soliton states can be explored at dissipative soliton transient region \cite{Ryczkowski2018}; the underlying seven different ultrafast phases are found in the Harmonic mode-locking build-up process \cite{Liu2019}; and the transient interference dynamics can be also observed in the build-up of the femtosecond mode-locking \cite{Herink2016}.

In this paper, the transient state of the RFL is characterized for the first time, and the experimental results are consistent well with the simulation results calculated by the generalized nonlinear Schr\"odinger equations. In the build-up region, the output power of the RFL grows up continuously fitting well with the Verhulst logistic model that is widely analyzed in biological growth dynamics, and the build-up duration is inversely related to the pump power. Besides, the spectral evolution of the RFL build-up is measured by chopping different RFL stages using acousto-optic modulators (AOMs), and we find it undergoes two phases from spectral density increase to spectral broadening. In the RFL dissipation region, the generated Stokes laser  dissipates immediately after switching off the pump power showing a high extinction ratio, and the remaining Stokes lightwave will all radiate out of the fiber after one round-trip. Our investigations pave a way to understand RFL lightwave kinetics at transient state. 

\section{Methods}

\subsection{Experimental setup}

\begin{figure}[ht!]
	\centering
	\includegraphics[width=14.5cm]{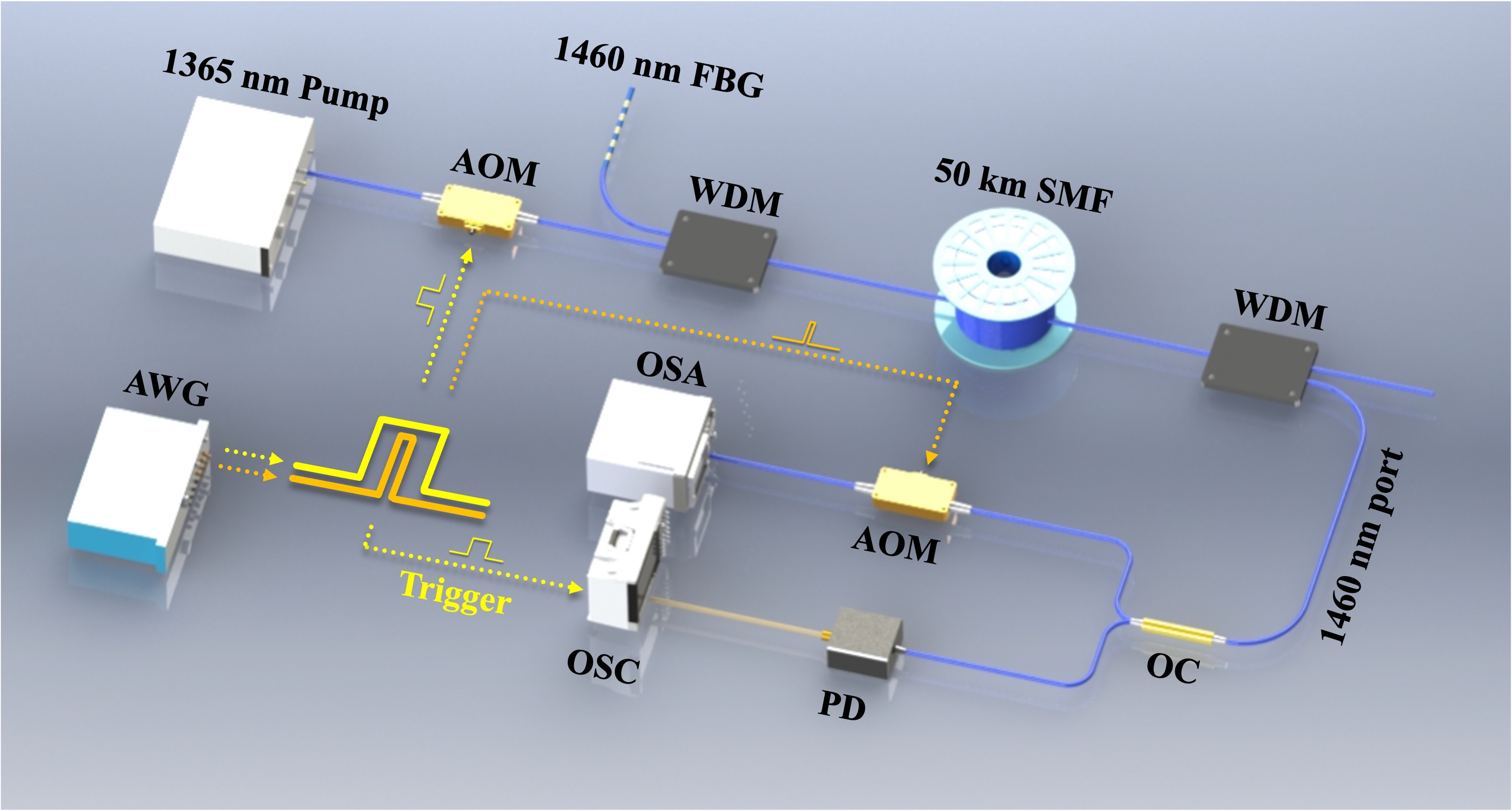}
	\caption{Experimental setup for investigating RFL transient state. }
	\label{fig1}
\end{figure}

The transient region of the RFL is investigated based on a forward-pumped RFL, as illustrated in Figure \ref{fig1}. The RFL is pumped by a 1365 nm Raman fiber laser and  switching of the pump is controlled by the first fiber acoustic-optic modulator (AOM1) with 2.3 dB insertion loss, 55 dB extinction ratio and about 35 ns rise/fall time for 1365 nm lightwave. The output pump is injected into a 50 km single mode fiber (SMF) through a 1365 nm/1460 nm wavelength division multiplexer (WDM), and the 1460 nm port is connected a 1460 nm fiber Bragg grating (FBG) playing the role of point reflector. The generated RFL outputs at the end of fiber, and is separated from the pump by an additional 1365 nm/1460 nm WDM. 

The temporal evolution of the RFL is detected by a 10 MHz photoreceiver and sampled by an oscilloscope (OSC) at 20 MSa/s sampling rate. For the spectral evolution of RFL, it is measured based on the time-slicing method \cite{Suret2013}, and a second fiber AOM (AOM2) is used in the scheme to realize 0.15 ms optical gate window that is shorter than the optical round-trip time. By varying the delay time between two AOMs' drive signals, we can record RFL spectral properties in arbitrary states by an optical spectrum analyzer (OSA, YOKOGAWA AQ6370D in this work).

\subsection{Simulation model}
Based on the experimental setup, we also performed simulations to theoretically analyzing the RFL transient state, including RFL build-up and dissipation. The simulation model is based on the generalized nonlinear Schr\"odinger equations (NLSEs) that can well describe the spectral and temporal dynamics of RFL \cite{Smirnov2013}:  

\begin{equation}
	\frac{\partial u_{p}^{\pm }}{\partial z}-\frac{1}{{{v}_{gs}}}\frac{\partial u_{p}^{\pm }}{\partial t}+i\frac{{{\beta}_{2p}}}{2}\frac{{{\partial }^{2}}u_{p}^{\pm }}{\partial {{t}^{2}}}+\frac{{{\alpha }_{p}}}{2}u_{p}^{\pm } = i{{\gamma }_{p}}{{\left| u_{p}^{\pm } \right|}^{2}}u_{p}^{\pm }-\frac{{{g}_{p}}\left( \omega  \right)}{2}\left( \left\langle {{\left| u_{s}^{\pm } \right|}^{2}} \right\rangle +\left\langle {{\left| u_{s}^{\mp } \right|}^{2}} \right\rangle  \right)u_{p}^{\pm },
	\label{eq1}
\end{equation}

\begin{equation}
	\frac{\partial u_{s}^{\pm }}{\partial z}+i\frac{{{\beta }_{2s}}}{2}\frac{{{\partial }^{2}}u_{s}^{\pm }}{\partial {{t}^{2}}}+\frac{{{\alpha }_{s}}}{2}u_{s}^{\pm }-\frac{\varepsilon \left( \omega  \right)}{2}u_{s}^{\mp }= i{{\gamma }_{s}}{{\left| u_{s}^{\pm } \right|}^{2}}u_{s}^{\pm }+\frac{{{g}_{s}}\left( \omega  \right)}{2}\left( \left\langle {{\left| u_{p}^{\pm } \right|}^{2}} \right\rangle +\left\langle {{\left| u_{p}^{\mp } \right|}^{2}} \right\rangle  \right)u_{s}^{\pm },
	\label{eq2}
\end{equation}
where $u$ is the complex envelope of the lightwave; $'p'$ and $'s'$ represent the pump and Stokes waves, respectively; '$+$' and '$-$' are the forward and backward propagating waves, respectively; $v_{gs}$ is the inverse group velocities between the pump and Stokes waves; $\omega$ is the angular frequency of lightwave; $\alpha$, $\gamma$,  $\beta_2$ , $\varepsilon$ and $g$  are the linear fiber loss, Kerr coefficient, second-order dispersion, Rayleigh scattering  and Raman gain, respectively. They are calculated and experimental measured in \cite{Lin2021} and \cite{Lin2022}. The total length of the fiber is defined as $L$.

\begin{table}[ht!]
	\footnotesize
	\caption{Parameters set in the simulation.}
	\label{table1}
	\tabcolsep 49pt
		\begin{tabular*}{\textwidth}{ccc}
			\toprule
			Parameter & Pump & Stokes \\
			\hline
			$\lambda$    & $1365\ nm$   & $1460\ nm$     \\
			$v_{g}$    &  $2.0509\times 10^{8}\ m/s$  & $2.0504\times 10^{8}\ m/s$  \\
			$\alpha$     & $0.30\ dB/km$        & $0.23\ dB/km$      \\
			Gain     & $5.3\times10^{-4}\ m^{-1}W^{-1}$ &   ---             \\
			$\varepsilon$  & $1\times10^{-7}\ m^{-1}$ & $0.6\times10^{-7}\ m^{-1}$ \\
			$\gamma$    & $0.0018\ m^{-1}W^{-1}$    &$0.0015\ m^{-1}W^{-1}$ \\
			$\beta_2$     & $8.3011\times10^{-27}\ s^{2}/m$ & $1.7972\times10^{-26}\ s^{2}/m$ \\
			$R_{R}$ & $4\times10^{-5}$  & $4\times10^{-5}$   \\
			$R_{L}$  & $4\times10^{-5}$    & $0.99$  \\
			\bottomrule
		\end{tabular*}
\end{table}

\begin{figure}[!t]
	\centering
	\includegraphics[width=10cm]{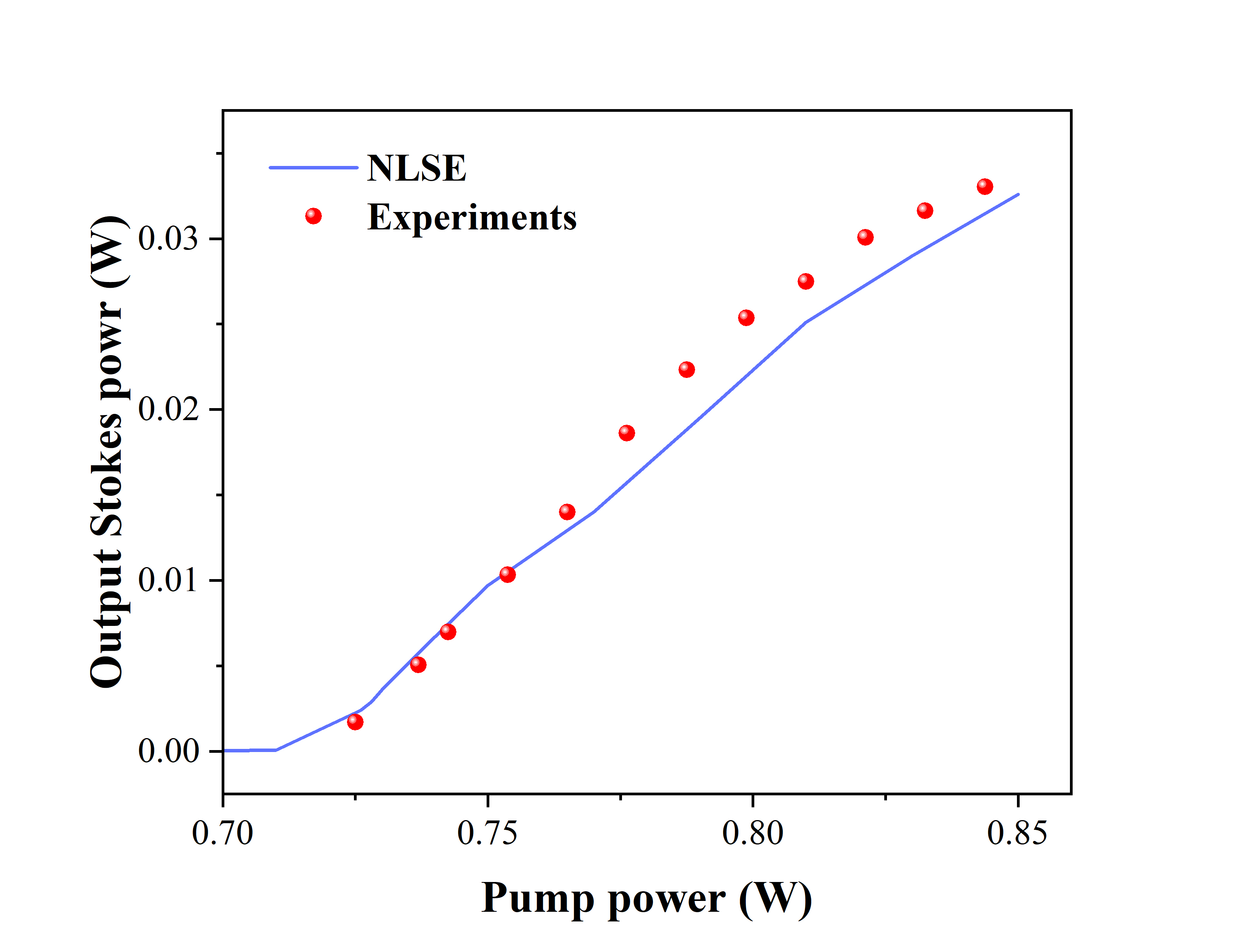}
	\caption{Output power of the RFL versus pump power (red dots: experimental data; solid line: numerical simulation results based on the NLSE model).}
	\label{fig2}
\end{figure}

In addition, the boundary conditions are $P_{p}^{+}(0,\omega,t)=P_{in}\left(\omega\right) T_{Lp}+R_{Lp}\left(\omega\right)P_{p}^{-}(0,\omega,t)$, $P_{p}^{-}(L,\omega,t)= R_{Rp}\left(\omega\right)P_{p}^{+}(L,\omega,t)$,   $P_{s}^{+}(0,\omega,t)= R_{Ls}\left(\omega\right)P_{s}^{-}(0,\omega, t)$ and $P_{s}^{-}(L,\omega, t)= R_{R s}\left(\omega\right)P_{s}^{+}(L,\omega,t)$, where $R_{L}\left(\omega\right)$ and $R_{R}\left(\omega\right)$ are wavelength-dependent reflectivity at left and right fiber ends respectively, and  $T_{L}$ is the corresponding transmittance. In addition, the value of $R_{Ls}(\omega)$ is experimentally measured with 0.2 nm bandwidth and 99\% reflectivity. Other parameter values in the simulation are summarized in Table \ref{table1}. The simulation is iterated by the split-step Fourier method \cite{Agrawal2006}, and the time-consuming Fourier transform process is performed in the GPU using CUDA to speed up the computation.

Figure \ref{fig2} shows the output Stokes power at the fiber end as a function of the pump power. The threshold of RFL is about 0.72 W in the experimental results. Above the threshold, the power of the Stokes laser increases rapidly, and its slope efficiency gradually slows down towards a specific value. It can be found that the simulation results are in accordance with the experimental results, validating the effectiveness of our simulation.

\section{Results and Discussion}

\begin{figure}[ht!]
	\centering
	\includegraphics[width=15cm]{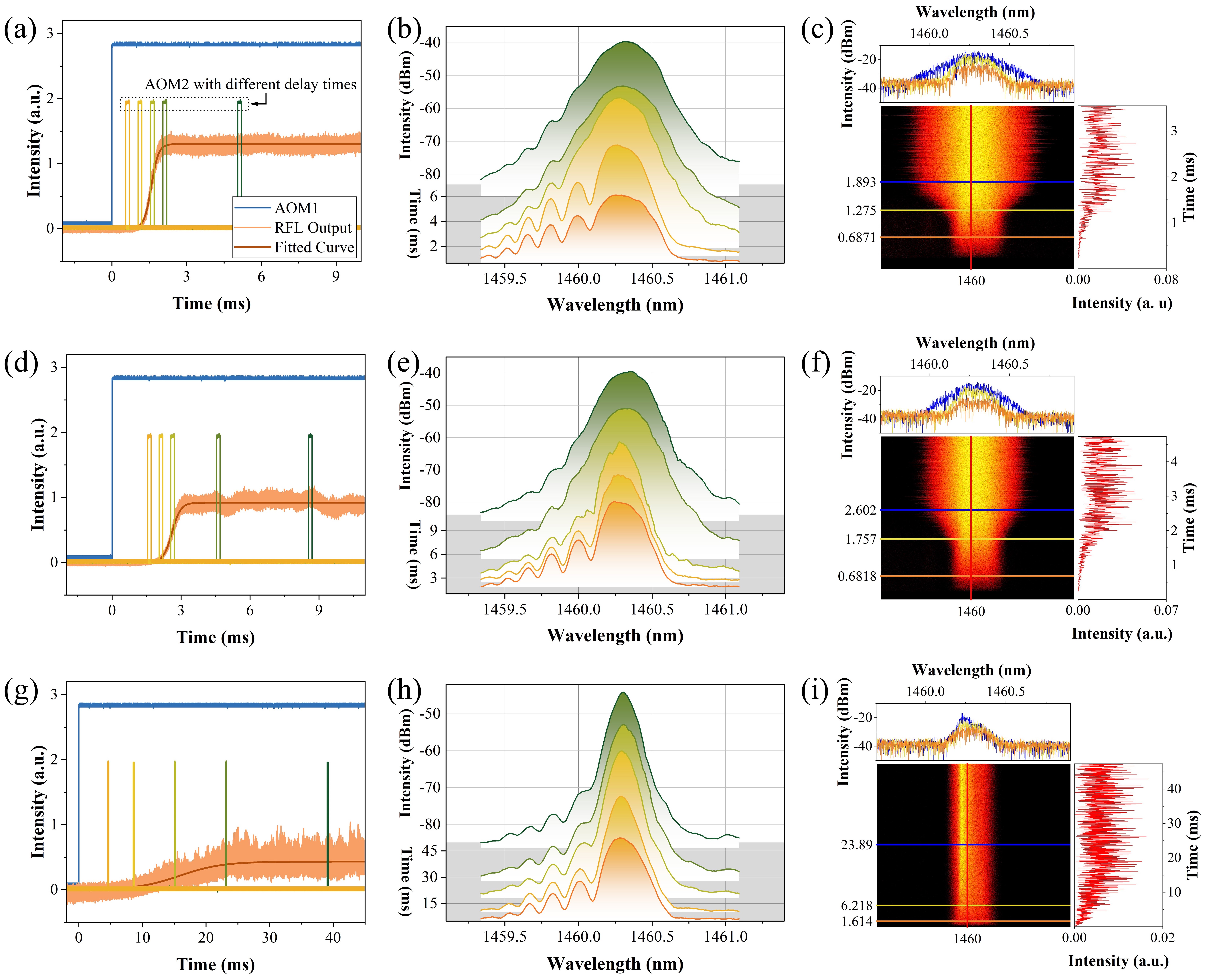}
	\caption{Temporal and spectral evolution of the RFL: (a)-(c) well above the threshold; (d)-(f) above the threshold; (g)-(i) around the threshold. (a), (d) and (g) temporal evolution in the experiment; (b), (e) and (h) spectral evolution in the experiment; (c), (f) and (i) temporal and spectral evolution in the simulation.}
	\label{fig3}
\end{figure}

Here, the transient state of RFL during the build-up process is first investigated. Figure \ref{fig3} shows the temporal and spectral evolution of the RFL when the pump power is suddenly switched on. The left-hand side and middle of figure \ref{fig3} are experimental results, and the right-hand side of  figure \ref{fig3} is corresponding simulation results.

For the RFL well above the threshold (0.85 W pump), temporal characteristics are described in figure \ref{fig3}(a), and the output power of the generated Stokes wave grows up continuously without discrete steps determined by the cavity round-trip time. The corresponding build-up time is about 2 ms, requiring only 4 optical round-trip time. The simulated temporal results shown in the red line spectral slice in figure \ref{fig3}(c) are consistent with experimental results.  For the spectral evolution of the RFL, it is measured based on the time-slicing method, and results are illustrated in figure \ref{fig3}(b). By varying the delay time between two AOMs' drive signals, we extracted 5 typical time slices in figure \ref{fig3}(a), covering some typical stages from the build-up of the RFL to the steady state. Line colors in figure \ref{fig3}(b) are consistent with the time-slicing extracted in figure \ref{fig3}(a). The spectral evolution undergoes two stages, and it can be observed in detail in the simulation (figure \ref{fig3}(c)). Initially, as the Stokes power grows up, the spectral density increases rapidly and the spectral bandwidth tends to narrow (from orange line time slice in figure \ref{fig3}(c) to yellow line). Further increase in spectral density makes the effect of the Kerr nonlinear gradually appear, resulting in RFL broadening (from yellow line time slice in figure \ref{fig3} (c) to blue line). 

Figure \ref{fig3} (d-f) and (g-i) are the evolution of the RFL above and around the threshold with 0.8 W and 0.73 W pump power, respectively. For the RFL above the threshold, the evolution of the RFL is similar to the phenomenon well above the threshold, but the laser build-up time is longer with about 4 ms corresponding to the 8 optical round-trip time. For the RFL around the threshold, the spectral broadening stage is not evident because the lower laser power is insufficient to induce significant Kerr effects. The build-up time for the RFL around the threshold increases dramatically to 20 ms, and the higher noise floor compared with figure \ref{fig3}(a) and (d) is caused by the larger gain chosen in the photoreceiver. 

It is worth noting that, regardless of the pump power, the growth curve of the Stokes laser is continuous,  indicating that the random distributed feedback of the Rayleigh scattering plays an important role in the generation of RFL. However, for the conventional  Raman fiber lasers, the laser is generated from the resonance of the cavity, thus the growth curve is step-wise in relation to the fiber cavity length \cite{Suret2013}, which is completely different from the RFL. 

In addition, the kinetics of RFL can be analogized to population dynamics in biology: Stokes photons are "predators" and pump photons are "prey". More "predators" will lead to fewer pump photons, and conversely, more "prey" will lead to more Stokes photons. The numbers of Stokes photons and pump photons will eventually reach equilibrium.  Here, we use Verhulst logistic growth model in biology domain \cite{Tsoularis2002} to describe the growth process of Stokes photons, which can be expressed as:

\begin{equation}
	{P}_{Sout}(t)=\frac{{{P}_{0}}P_{max}}{{{P}_{0}}+\left( P_{max}-{{P}_{0}} \right){{G}^{-t}}},
	\label{eq3}
\end{equation}
where $P_{Sout}$ is the Stokes output power; $P_{0}$ is the initial noise from the amplifier spontaneous emission; $P_{max}$ is the maximum output power; $G$ is the growth rate of the Stokes power. As illustrated in left-hand side of figure \ref{fig3}, the profile of RFL power growth is fitted well with Verhulst logistic growth model. RFL growth rate $G$ well above the threshold is as high as 1030 V/s, will drop to  226 $V/s$ above the threshold, and even as low as 1.3 $V/s$ at the threshold.  Thus, by varying the pump power, we can use random lasers to analyze biological population dynamics with different growth rate. Based on the RFL system, it may open up new important avenues for the understanding of more biological phenomena.

\begin{figure}[ht!]
	\centering
	\includegraphics[width=10cm]{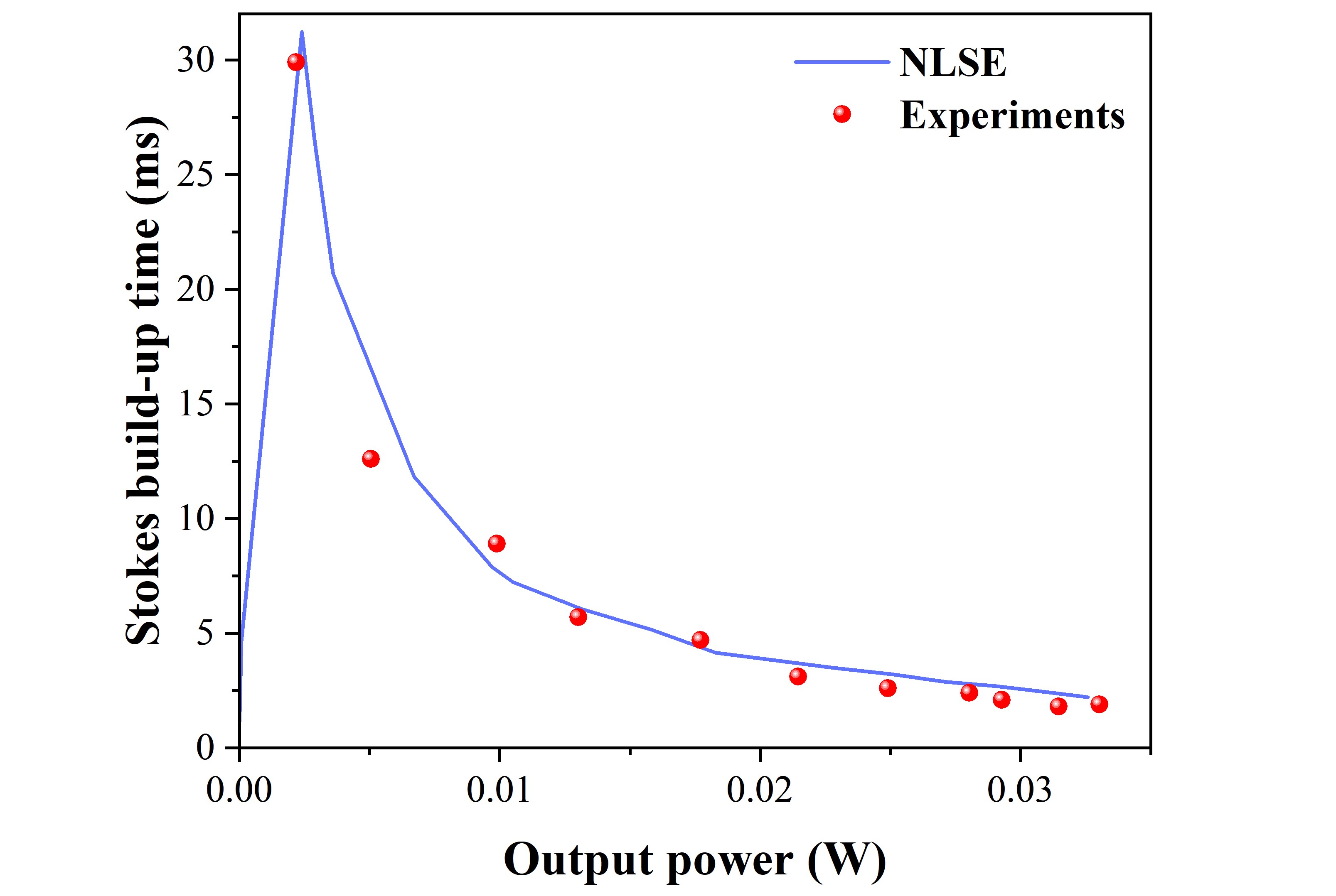}
	\caption{The build-up time of the forward-pumped RFL at different output Stokes powers.}
	\label{fig6}
\end{figure}

For the threshold, RFL exhibits extremely complex physical processes, including a sudden narrowing of the laser linewidth \cite{Kolokolov2015}, L\'evy statistical behavior \cite{Li2019}, ultra-narrow spectral modes dynamics \cite{Kirik2021}. Here, we also find the build-up time of the RFL takes up abruptly around the threshold from 3 ms (build-up time of amplifier spontaneous emission) to 30 ms, as illustrated in figure \ref{fig6}. Above the threshold, the build-up time of the RFL is inversely related to the Stokes output power. With further increasing the pump power, the Stokes build-up time gradually decreases to 10 ms at 0.01 W output power, and then to 2 ms at 0.03 W output power.  The results in figure \ref{fig6} are important for the RFL point sensing \cite{Wang2012, Lin2022}. When the center wavelength of the sensing FBG is shifted caused by the external disturbance, it takes time to re-establish the RFL to deliver the sensing information, which limits the RFL sensing bandwidth. The duration from pump switch-on to the steady-state in this work is the longest laser re-establish time that determines the lower bound of the sensing bandwidth. Thus, based on the results of figure \ref{fig6},  we can appropriately increase the pump power for enhancing the sensing bandwidth.

\begin{figure}[ht!]
	\centering
	\includegraphics[width=14.5cm]{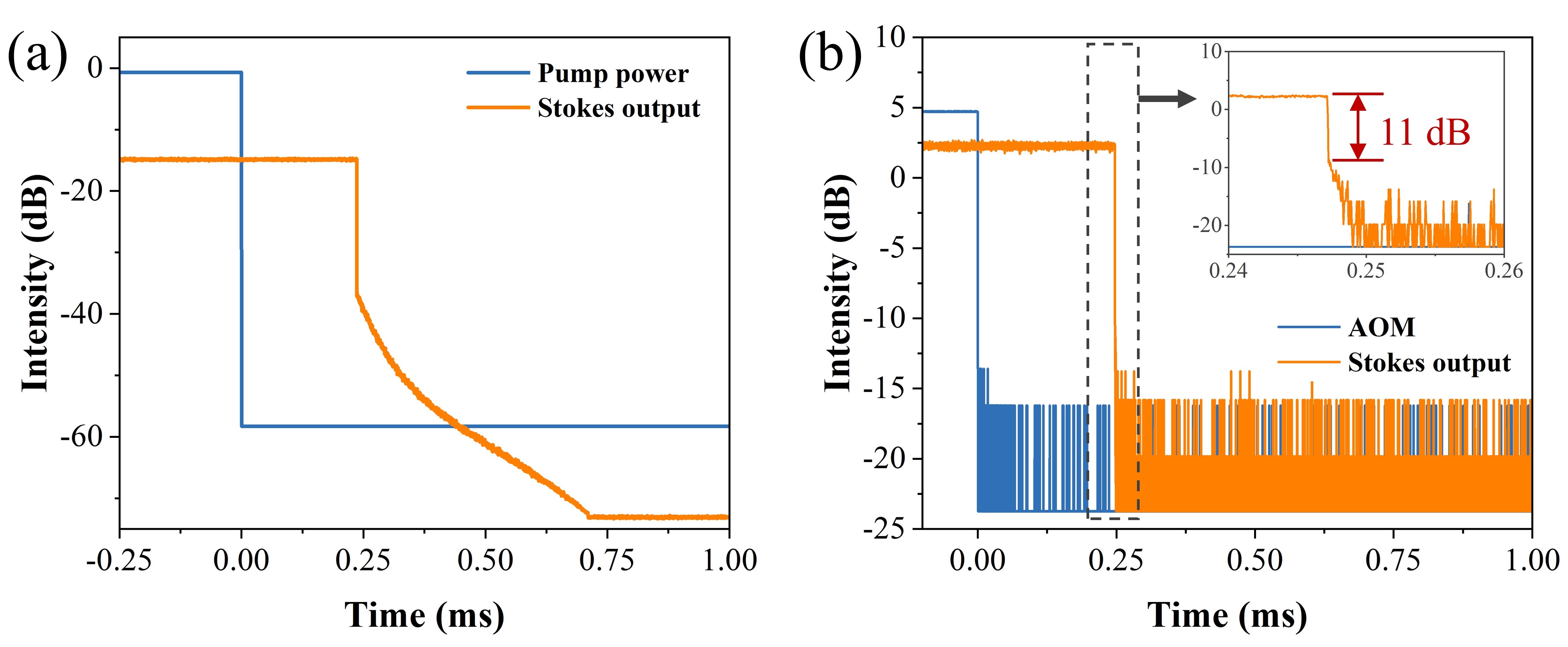}
	\caption{Temporal evolution of the RFL after switch-off the pump.(a) simulation results; (b) experimental results.}
	\label{fig7}
\end{figure}

The transient state of RFL during the pump switch-off state is also investigated. When the pump power is switched off, the generated Stokes laser will drop off suddenly after half of the round-trip time, and the remaining Stokes lightwave from the previous Rayleigh scattering will all radiate out of the fiber after one round-trip, as shown in Figure \ref{fig7} (a). Figure \ref{fig7}(b) are the experiment results after switch-off the pump, and the phenomenon is similar to the simulation results showing exponential decay. The decay process is recorded in several microseconds, much higher than the rise/fall time of the photoreceiver ($ \sim $ 80 ns). And then, the Stokes signal is drowned out in the measuring noise. It is worth noting that the extinction ratio of RFL falling edge is as high as $\sim$ 11 dB, showing the ability to generate Q-switched  pulse with high quality.

\section{Conclusion}
In this work, we analyze the transient state of the RFL for the first time. In the RFL build-up region,  the temporal evolution of RFL shows Verhulst logistic growth curves and the build-up time is inversely related to the pump power. Besides, the spectral evolution of the RFL contains two stages from spectral density increase to spectral broadening. On the contrary, after switching off the pump, the generated Stokes laser dissipates immediately, then the remaining Stokes lightwave will gradually radiate out of the fiber. This work provides a new window into the underlying complex physics of the RFL dynamics, and the results would be also beneficial for researching other complex systems, like biological dynamics, rogue-wave build-up, etc.

\newpage

\Acknowledgements{This work is supported by the Natural Science Foundation of China (62075030), National Ten-Thousand Talent Program (W030211001001), and Sichuan Provincial Project for Outstanding Young Scholars in Science and Technology (2020JDJQ0024).
}


\bibliographystyle{IEEEtran}
\bibliography{MSP-template}



\end{document}